\documentclass[aps,prl,twocolumn,preprintnumbers,10pt,showpacs,nofootinbib]{revtex4-1}

\usepackage{amsmath,amsfonts,epsfig,color,latexsym}
\usepackage{amssymb}
\usepackage[english, USenglish]{babel}
\usepackage{epsfig,psfrag}
\usepackage{slashed}
\usepackage[normalem]{ulem}
\usepackage{color}

\usepackage{soul,xcolor}
\usepackage{hyperref}
\setstcolor{red}
\usepackage{comment}
\usepackage[utf8]{inputenc}

\begin{document}


\newcommand{\norm}[1]{{\protect\normalsize{#1}}}
\newcommand{\p}[1]{(\ref{#1})}
\newcommand{\half}{\tfrac{1}{2}}
\newcommand \vev [1] {\langle{#1}\rangle}
\newcommand \ket [1] {|{#1}\rangle}
\newcommand \bra [1] {\langle {#1}|}
\newcommand \pd [1] {\frac{\pa}{\pa {#1}}}
\newcommand \ppd [2] {\frac{\pa^2}{\pa {#1} \pa{#2}}}
\newcommand{\ed}[1]{{\color{red} {#1}}}

\newcommand{\cI}{{\cal I}}
\newcommand{\cM}{{\cal M}}
\newcommand{\cR}{{\cal R}}
\newcommand{\cS}{{\cal S}}
\newcommand{\cK}{{\cal K}}
\newcommand{\cL}{{\cal L}}
\newcommand{\cF}{{\cal F}}
\newcommand{\cN}{{\cal N}}
\newcommand{\cA}{{\cal A}}
\newcommand{\cB}{{\cal B}}
\newcommand{\cG}{{\cal G}}
\newcommand{\cO}{{\cal O}}
\newcommand{\cY}{{\cal Y}}
\newcommand{\cX}{{\cal X}}
\newcommand{\cT}{{\cal T}}
\newcommand{\cW}{{\cal W}}
\newcommand{\cP}{{\cal P}}
\newcommand{\bP}{{\bar\Phi}}
\newcommand{\mK}{{\mathbb K}}
\newcommand{\nt}{\notag\\}
\newcommand{\pa}{\partial}
\newcommand{\ep}{\epsilon}
\newcommand{\om}{\omega}
\newcommand{\bom}{\bar\omega}
\newcommand{\etap}{\bar\epsilon}
\newcommand{\vep}{\varepsilon}
\renewcommand{\a}{\alpha}
\renewcommand{\b}{\beta}
\newcommand{\g}{\gamma}
\newcommand{\s}{\sigma}
\newcommand{\la}{\lambda}
\newcommand{\tl}{\tilde\lambda}
\newcommand{\tm}{\tilde\mu}
\newcommand{\tk}{\tilde k}
\newcommand{\da}{{\dot\alpha}}
\newcommand{\db}{{\dot\beta}}
\newcommand{\dg}{{\dot\gamma}}
\newcommand{\dd}{{\dot\delta}}
\newcommand{\q}{\theta}
\newcommand{\bq}{{\bar\theta}}
\renewcommand{\r}{\rho}
\newcommand{\br}{\bar\rho}
\newcommand{\be}{\bar\eta}
\newcommand{\bQ}{\bar Q}
\newcommand{\bx}{\bar \xi}
\newcommand{\tx}{\tilde{x}}
\newcommand{\tr}{\mbox{tr}}
\newcommand{\+}{{\dt+}}
\renewcommand{\-}{{\dt-}}
\newcommand{\ti}{{\textup{i}}}

\newcommand{\dlog}{d{\rm log}}
\newcommand{\tred}[1]{\textcolor{red}{\bfseries #1}}
\newcommand{\eps}{\epsilon}
\newcommand{\sectioninline}[1]{{\bfseries #1}}

\newcommand\clim[2]{\overset{#1||#2}{\rightarrow}}
\newcommand\climj[2]{{#1||#2}}

\preprint{IPPP/19/37, MITP/19-031, MPP-2019-88, USTC-ICTS-19-12, ZU-TH 23/19}

\title{Analytic form of the full
 two-loop five-gluon all-plus helicity amplitude
}

\author{S.\ Badger$^{a}$, D.\ Chicherin$^{b}$, T.\ Gehrmann$^{c}$, G.\ Heinrich$^{b}$, J.\ M.\ Henn$^{b}$, T.\ Peraro$^{c}$, P.\ Wasser$^{d}$, Y.\ Zhang$^{b,e}$, S.\ Zoia$^{b}$}

\affiliation{
$^a$ Institute for Particle Physics Phenomenology, Durham University, Durham DH1 3LE, United Kingdom\\
$^b$ Max-Planck-Institut f{\"u}r Physik, Werner-Heisenberg-Institut, D-80805 M{\"u}nchen, Germany\\
$^c$ Physik-Institut, Universit{\"a}t Z{\"u}rich, Wintherturerstrasse 190, CH-8057 Z{\"u}rich, Switzerland\\
$^d$ PRISMA+ Cluster of Excellence, Johannes Gutenberg University, D-55099 Mainz, Germany\\
$^e$ Interdisciplinary Center for Theoretical Study, University of Science and Technology of China, Hefei, Anhui 230026, China
}

\begin{abstract}
We compute the full-color two-loop five-gluon amplitude for the all-plus helicity configuration.
In order to achieve this, we calculate the required master integrals for all permutations of the external legs,
in the physical scattering region.
We verify the expected divergence structure of the amplitude, and extract the finite hard function.
We further validate our result by checking the factorization properties in the collinear limit.
Our result is fully analytic and valid in the physical scattering region.
We express it in a compact form containing logarithms, dilogarithms and rational functions.
\end{abstract}

\maketitle

The abundant amount of data to be collected by the ATLAS and CMS
collaborations in future runs of the Large Hadron Collider at CERN opens up
 a new era of precision physics. Some of the most prominent precision observables are related to 
 three-jet production~\cite{Aaboud:2017fml,Sirunyan:2018adt}, which allows in-depth studies of the
 strong interaction up to the highest energy scales, including precision measurements of the
 QCD coupling constant $\alpha_s$ and its scale evolution. The physics exploitation
 of these precision data requires highly accurate theory predictions, which are obtained
 through the computation of higher orders in perturbation theory. Second-order
 corrections (next-to-next-to-leading order, NNLO) were computed recently for many
 two-to-two scattering processes, including two-jet production~\cite{Currie:2017eqf}.
A comparable level of theoretical accuracy could up to now not be obtained for
genuine two-to-three processes, especially since the relevant matrix elements
for processes involving five external partons including full color are known only up to one loop~\cite{Bern:1993mq,Bern:1994fz,Kunszt:1994nq}.

The evaluation of these two-loop five parton matrix elements faces two types of
challenges: to relate the large number of two-loop integrands to a smaller number of master
integrals, and to compute these master integrals (two-loop five-point functions).
Important progress was made most recently on both issues, with the development
and application of efficient integral reduction techniques, either
analytical~\cite{vonManteuffel:2014ixa,Larsen:2015ped,Peraro:2016wsq,Badger:2017jhb,Boehm:2017wjc,Boehm:2018fpv,Kosower:2018obg,Chawdhry:2018awn}
 or semi-numerical~\cite{Ita:2015tya,Abreu:2017idw},
as well as with the computation of the two-loop five-point functions for
planar~\cite{Gehrmann:2015bfy,Papadopoulos:2015jft,Gehrmann:2018yef} and
non-planar~\cite{Chicherin:2018mue,Chicherin:2018old}
integral topologies. The latter developments already have led to first results for two-loop five-point amplitudes in
supersymmetric Yang-Mills theory~\cite{Abreu:2018aqd,Chicherin:2018yne} and supergravity~\cite{Chicherin:2019xeg,Abreu:2019rpt}.

The recent progress enabled the computation of the full set of the leading-color
two-loop corrections to the five-parton amplitudes, represented in a
semi-numerical form~\cite{Badger:2013gxa,Badger:2017jhb,Abreu:2017hqn,Abreu:2018jgq}.
These results are establishing the technical methodology, their evaluation is
however too inefficient for practical use in the computation of collider cross sections.
Towards this aim, analytic results are preferable, which were obtained so
far only at leading color for the five-parton amplitudes~\cite{Gehrmann:2015bfy,Badger:2018enw,Abreu:2018zmy,Abreu:2019odu}.
Besides the more efficient numerical evaluation, these results also
allow for detailed investigations of the limiting behavior in kinematical
limits, thereby elucidating the analytic properties of scattering in QCD.

The leading-color corrections consist only of planar Feynman diagrams. At subleading color
level, non-planar diagrams and integrals contribute as well, leading to
a considerable increase in complexity, both in the reduction of the integrand and in the
evaluation of the master integrals. In this Letter, we make the first step towards
the fully analytic evaluation of two-loop five-point amplitudes, by exploiting the
recently derived non-planar two-loop five-point master integrals~\cite{Chicherin:2018mue,Chicherin:2018old} to obtain
an analytic expression for the two-loop five-gluon amplitude with all-plus helicities~\cite{Badger:2015lda}.

\sectioninline{Kinematics.}
We study the scattering of five gluons in the all-plus helicity configuration.
The corresponding amplitude has a complete permutation symmetry under exchange of external gluons.
The five light-like momenta $p_{i}$ are subject to on-shell and momentum conservation conditions,
$p_{i}^2= 0$, and $\sum_{i=1}^{5} p_i = 0$, respectively. They give rise to the following independent parity-even Lorentz invariants
\begin{align}\label{kinematicvariables}
X = \{ s_{12} \,,  s_{23} \,,  s_{34} \,, s_{45} \,, s_{15} \} \,,
\end{align}
with $s_{ij} = 2 \, p_{i} \cdot p_{j}$,
as well as to the parity-odd invariant \mbox{$\eps_5 = \text{tr}(\gamma_5 \slashed{p}_1 \slashed{p}_2 \slashed{p}_3 \slashed{p}_4)$}. The latter is related to the Gram determinant $\Delta = \text{det}(s_{ij}  |_{i,j=1}^4 )$ through $\eps_5^2 = \Delta$.

Without loss of generality, we take the kinematics to lie in the $s_{12}$ scattering region.
The latter is defined by all $s$-channel invariants being positive, i.e.
\begin{align}
\label{eq:s-channel}
s_{12}>0 \, , \qquad s_{34}>0 \, , \qquad s_{35}>0 \, , \qquad s_{45}>0 \, ,
\end{align}
and $t$-channel ones being negative, i.e.
\begin{align}
\label{eq:t-channel}
s_{1j}<0 \, , \   s_{2j}<0 \, , \quad {\rm for} \; j=3,4,5\,,
\end{align}
as well as by the requirement that
the particle momenta are real, which implies $\Delta<0$.

The external momenta $p_{i}$ lie in four-dimensional Minkowski space.
We will encounter $D-$dimensional Feynman integrals, with $D=4-2\eps$,
and the loop momenta therefore live in $D$ dimensions.
We keep the explicit dependence on the spin dimension $D_s = g^{\mu}_{\ \mu}$ of the gluon,
which enters the calculation via the integrand numerator algebra. {Results in the t'Hooft-Veltman \cite{tHooft:1972tcz} and Four-Dimensional-Helicity~\cite{Bern:2002zk} schemes can be obtained by setting $D_s=4-2 \epsilon$ and $D_s=4$, respectively.
We denote $\kappa = (D_s -2)/6$.}

\sectioninline{Decomposition of the amplitude in terms of color structures.}
We expand the unrenormalized amplitude in the coupling 
$a =g^2 \, e^{-\epsilon \, \gamma_{\text{E}}}/(4 \pi)^{2- \epsilon}$ as
\begin{align}\label{couplingexpansion}
\mathcal{A}_5 = i \, g^3 \, \sum_{\ell \ge 0}
  a^{\ell} \, \mathcal{A}^{(\ell)}_5 \, .
\end{align}
Due to the particular helicity-configuration, the amplitude vanishes at tree-level~\cite{Parke:1985pn,Mangano:1990by},
 and is hence finite at one loop.

The amplitude is a vector in color space.
Adopting the conventions of Ref.~\cite{Edison:2011ta}, we decompose the one- and two-loop amplitude as
\begin{align}
\label{eq:ColorDecomposition1loop}
& \mathcal{A}_5^{(1)} = \sum_{\lambda=1}^{12} N_c \, A_{\lambda}^{(1,0)}   \cT_{\lambda}+\sum_{\lambda=13}^{22} A_{\lambda}^{(1,1)}  \cT_{\lambda}\, , \\
\label{eq:ColorDecomposition}
& \mathcal{A}_5^{(2)} = \sum_{\lambda=1}^{12} \left(N_c^{2} A_{\lambda}^{(2,0)} +A_{\lambda}^{(2,2)} \right) \cT_{\lambda}+\sum_{\lambda=13}^{22}  N_c A_{\lambda}^{(2,1)}  \cT_{\lambda}\,.
\end{align}
Here the $\{ \cT_{\lambda} \}$ consist of 12 single traces, $\lambda=1,\ldots,12$,
and 10 double traces, $\lambda=13,\ldots,22$. We have
\begin{equation}
\begin{aligned}
& \cT_1 = \text{Tr}(12345)-\text{Tr}(15432)\, ,\\
& \cT_{13} = \text{Tr}(12)\left\lbrack \text{Tr}(345)-\text{Tr}(543)\right\rbrack\, , \label{tdef}
\end{aligned}
\end{equation}
where $\text{Tr}(i_1 i_2 \, ... \, i_n) \equiv \text{Tr}(T^{a_{i_1}} \, ... \, T^{a_{i_n}})$ denotes the trace of the generators $T^{a_i}$ of the fundamental representation of $SU(N_c)$.
The remaining color basis elements $\cT_{\lambda}$ are given by permutations of $\cT_1$ and $\cT_{13}$.
For the explicit expressions, see Eqs.~(2.1) and (2.2) of~\cite{Edison:2011ta}.

The one-loop expression can be found in \cite{Bern:1993mq}.
Here we write it in a new form,
\begin{align}\label{A1loop}
A_{1}^{(1,0)} =  \frac{\kappa}{5}  \,   \sum_{S_{\mathcal{T}_{1}}}  \,  \biggl[ \frac{[24]^2}{\vev{13} \vev{35} \vev{51}}  + 2 \, \frac{ [23]^2}{\vev{14} \vev{45} \vev{51}}  \biggr] \,,   
\end{align}
up to $\cO(\eps)$ terms.
The sum runs over the subset $S_{\mathcal{T}_{\lambda}}$ of permutations of the external legs that leave $\mathcal{T}_{\lambda}$ invariant.
All other terms in (\ref{eq:ColorDecomposition1loop}) follow from symmetry, and from $U(1)$ decoupling relations.

The new representation (\ref{A1loop}) makes a symmetry property manifest. The basic rational object is invariant under conformal transformations, which are defined as \cite{Witten:2003nn}
\begin{align}\label{Wittenconformal}
k_{\alpha \dot{\alpha}} = \sum_{i=1}^{5} \frac{\partial^2}{\partial \lambda^{\alpha}_i \partial \tilde\lambda^{\dot\alpha}_i}\,.
\end{align}
The property $k_{\alpha \dot{\alpha}} \mathcal{A}_{5}^{(1)} =  \cO(\eps)$   
is obvious term-by-term due to the form of the operator in Eq. (\ref{Wittenconformal}).

In this Letter, we compute the full two-loop amplitude. 
The leading color single trace terms $A_{\lambda}^{(2,0)}$ were computed in Refs. \cite{Gehrmann:2015bfy,Dunbar:2016aux}.
Generalizations of the $U(1)$ decoupling relation imply that the most subleading color terms $A_{\lambda}^{(2,2)}$ can be obtained from the leading single trace $A_{\lambda}^{(2,0)}$  and the double trace terms $A_{\lambda}^{(2,1)}$ ~\cite{Edison:2011ta}.
We present explicitly the result for the finite part of the double trace term $A_{13}^{(2,1)}$.
The other double trace terms can be obtained by analytic continuation, as explained below.

\sectioninline{Factorization and exponentiation of infrared divergences.}
\label{sec:IRfactorization}
Infrared divergences (soft and collinear) in loop amplitudes factorize similarly to ultraviolet divergences, in the following way:
\begin{align}
\mathcal{A} = \mathcal{Z} \, \mathcal{A}^{f}\,.
\end{align}
Here, the factor $\mathcal{Z}$ is a matrix in color space. It contains all infrared divergences, in the sense
that we can define an infrared finite hard function according to
\begin{align}
\mathcal{H} = \lim_{\eps \to 0} \mathcal{A}^{f}\,.
\end{align}
For massless scattering amplitudes, $\mathcal{Z}$ is known to three loops, see \cite{Catani:1998bh,Becher:2009cu,Gardi:2009qi}.
In the present case, the tree-level amplitude vanishes,
and we therefore need only the one-loop part of the infrared matrix,
$\mathcal{Z} = \mathbb{\textbf{1}} + a \, \mathcal{Z}^{(1)}$, with
\begin{align}
\label{eq:Zfactor}
\mathcal{Z}^{(1)} =
  - \frac{ e^{\epsilon \, \gamma_{\text{E}}}}{ \epsilon^2 \, \Gamma(1-\eps) } \sum_{i \neq j}^{5}  \, \vec{\mathbf{T}}_i \cdot \vec{\mathbf{T}}_j \left(\frac{\mu^2}{- s_{ij}} \right)^{\epsilon} \, ,
\end{align}
where $\mu$ is the dimensional regularization scale, and
$\vec{\mathbf{T}}_i = \{\mathbf{T}^a_i \}$ are the generators of $SU(N_c)$ in the adjoint representation of gluon $i$, with $\mathbf{T}_i^b \circ T^{a_i} = - i f^{b a_i c_i} T^{c_i}$.
We set $\mu = 1$. The explicit dependence can be recovered from dimensional analysis.

\begin{figure}[t]
  \begin{center}
    \includegraphics[width=0.32\columnwidth]{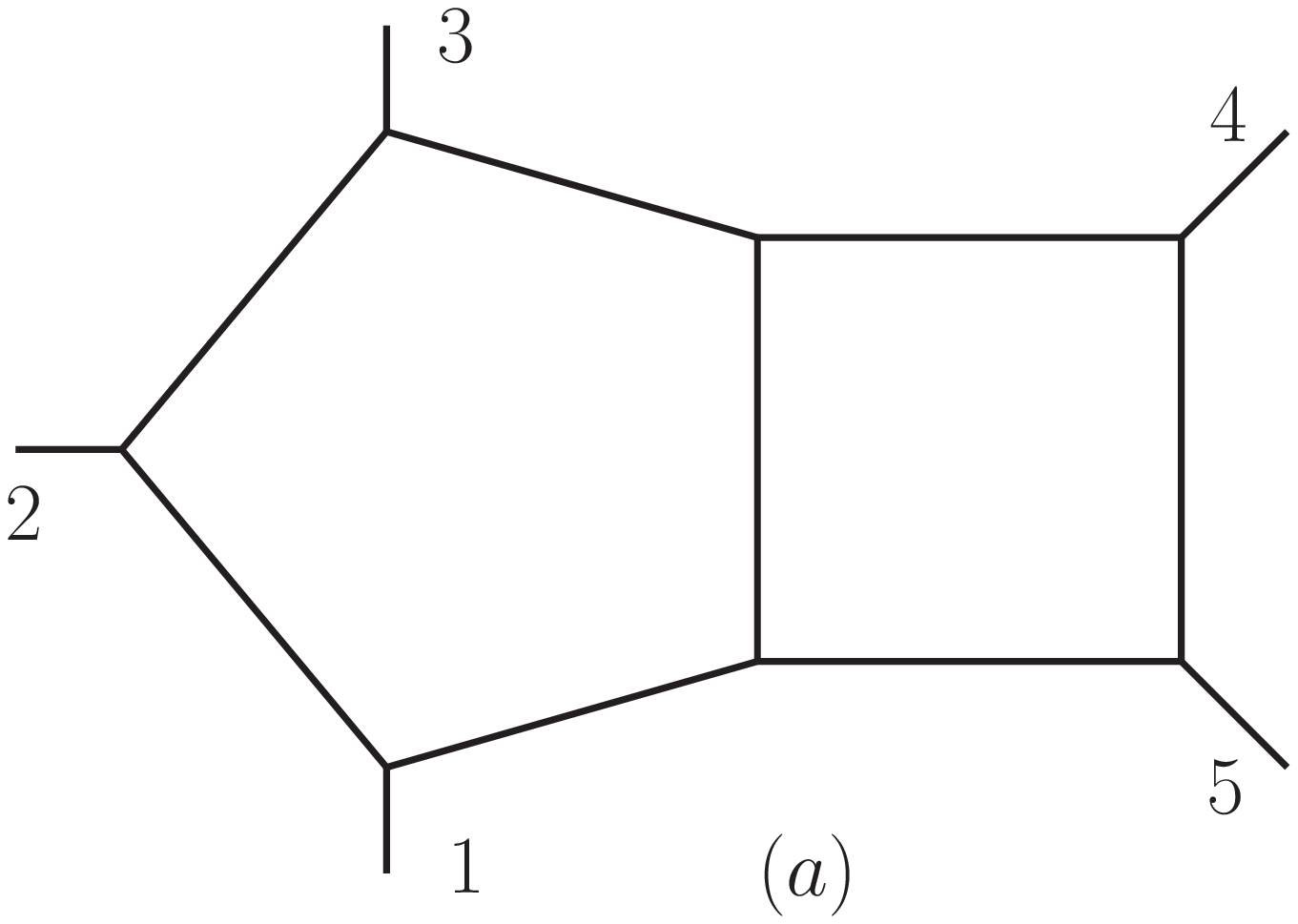}
    \includegraphics[width=0.32\columnwidth]{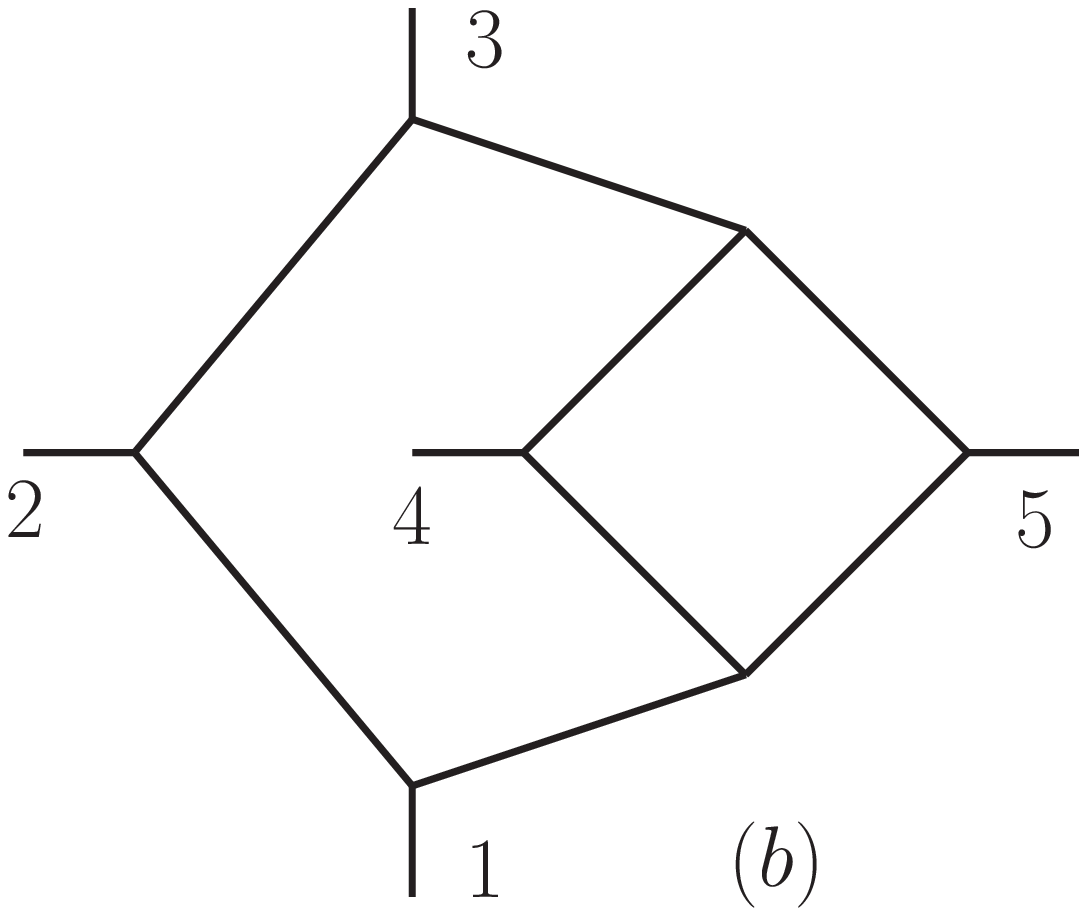}
\includegraphics[width=0.32\columnwidth]{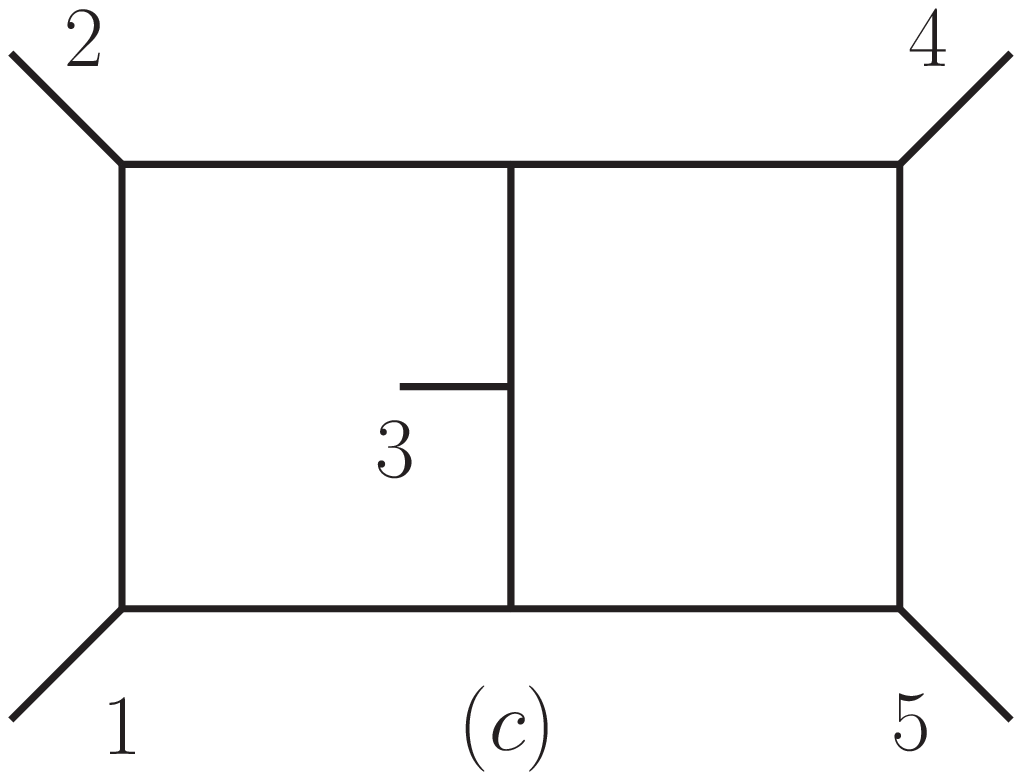}
    \caption{Two-loop five-particle Feynman integral topologies.}
    \label{fig:Feynman_integrals}
  \end{center}
\end{figure}

\sectioninline{Two-loop integrand and finite field reduction to uniform-weight master integrals.}
The starting point of our calculation is the integrand presented in \cite{Badger:2015lda}.
The latter was obtained using modern generalized unitarity techniques.
It is given in terms of the integrals of the type shown in Fig.~\ref{fig:Feynman_integrals} (and similar integrals corresponding to subgraphs), with certain numerator factors.
We begin by rewriting scalar products of $-2\, \epsilon$-dimensional components of loop momenta in terms of Gram determinants, see e.g. \cite{Chicherin:2018old}.
In this way, one obtains numerators with up to degree five for the eight-propagator integrals shown in
Fig.~\ref{fig:Feynman_integrals}, as well as some of degree six for the one-loop squared sectors.
This is significantly higher as compared to the previous calculations in $\mathcal{N}=4$ super Yang-Mills \cite{Abreu:2018aqd,Chicherin:2018yne,Chicherin:2018old} and $\mathcal{N}=8$ supergravity \cite{Chicherin:2019xeg,Abreu:2019rpt}, where numerators of up to degree one and two were required, respectively.

We set up a system of integration by parts (IBP) identities  with the help of \textsc{LiteRed}~\cite{Lee:2012cn}.
The task is to reduce the IBP system to a minimal required set of reduction identities.
This is a difficult problem, and to solve it we profit from novel
finite field and functional reconstruction techniques ~\cite{vonManteuffel:2014ixa,Peraro:2016wsq}.
To do this, we solve the system, modulo prime integers, for numerical (rational) values of $\epsilon$ and of the kinematic invariants $s_{ij}$, using a custom linear solver for sparse systems. In doing so, we use a basis of dlog \cite{ArkaniHamed:2010gh,Henn:2013pwa,Chicherin:2018old,Henn:2019rmi}
master integrals as preferred integrals during the solution. We then reconstruct the analytic results from these numerical evaluations using a multivariate reconstruction algorithm, based on the one described in Ref.~\cite{Peraro:2016wsq}.
The calculation is further significantly simplified by reconstructing reduction tables only for the relevant combinations of integrals appearing in the representation of the amplitude.

\sectioninline{Analytic results for master integrals in the $s_{12}$ scattering region.}
In previous work, the planar master integrals shown in Fig.~\ref{fig:Feynman_integrals}(a)
were computed in all kinematic scattering regions \cite{Gehrmann:2015bfy,Papadopoulos:2015jft,Gehrmann:2018yef};
the master integrals of Fig.~\ref{fig:Feynman_integrals}(b) and Fig.~\ref{fig:Feynman_integrals}(c) were computed
in one kinematic region only.

Since the integrals enter the amplitude in all different permutations of the external legs,
we need to know them in several kinematic regions.
In principle, the answer in different kinematic regions can be obtained via
analytic continuation, see \cite{Gehrmann:2018yef,Henn:2019rgj}.
Here we adopt a different strategy: we consider all permutations of all required master integrals,
together with the differential equations they satisfy,
and compute them directly in the $s_{12}$ channel.
Taking the permutations of the differential equations is unambiguous, as the differential
matrices are rational functions of the kinematics. In order to streamline the calculation, 
we also identify relations between integrals with permuted external legs.
In this way, we do not need to continue the functions analytically.
This workflow is also less error-prone, as all steps are completely automatic.

The dlog master integrals $\vec{f}$ of each family satisfy a differential equation of the form~\cite{Henn:2013pwa}
\begin{align}\label{DEX}
\partial_{X} \vec{f}(X,\eps) = \eps \, \partial_{X} \left[ \sum_{i=1}^{31} a_{i}  \log W_{i}(X)  \right]  \vec{f}(X,\eps) \, ,
\end{align}
where $a_i$ are constant matrices, and $W_i(X)$ are letters of the so-called pentagon alphabet~\cite{Chicherin:2017dob}, algebraic functions of the kinematic variables $X$ encoding the branch cut structure of the solution. The matrices $a_i$ are peculiar to the family and to the precise choice of basis $\vec{f}$, but the set of letters $\{ W_i \}$ is the same for all massless two-loop five-particle integrals.

Solving the differential equations requires a boundary point. We choose
\begin{align}
X_{0} = \{ 3 , -1, 1, 1,-1 \}\,.
\end{align}
This point lies in the $s_{12}$ scattering region, and is symmetric under $p_{1} \leftrightarrow p_{2}$, or any permutation of $\{p_3, p_4 , p_5 \}$.
We fix the boundary values analytically by requiring the absence of unphysical singularities.
See Refs.~\cite{Gehrmann:2018yef,Chicherin:2018mue,Henn:2019rgj} for a more detailed discussion.
In this way, all boundary values are related to a few simple integrals. The latter are found in the literature \cite{Gehrmann:2000zt,Gehrmann:2001ck}.

We verify the boundary values numerically for two permutations of the integrals of Fig.~\ref{fig:Feynman_integrals}(c),
and for some integrals of Fig.~\ref{fig:Feynman_integrals}(b).
This is done by computing all master integrals numerically, using {\sc pySecDec}~\cite{Borowka:2017idc, Borowka:2018goh},
at the symmetric point $X_{0}$.
We find it convenient to do this for an integral basis in $D=6-2\eps$ dimensions.
Moreover, we check the boundary values of the planar integrals of Fig.~\ref{fig:Feynman_integrals}(a) against the program provided with Ref.~\cite{Gehrmann:2018yef}.

We expand the solution to (\ref{DEX}) in $\eps$ up to order $\eps^4$, corresponding to weight four functions.
The latter are expressed in terms of Chen iterated integrals. We adopt the same notation as in Ref. \cite{Gehrmann:2018yef}, and write the iterated integrals as
\begin{align}
[W_{i_1},\ldots , W_{i_n}]_{X_0}(X) & =  \int_{\gamma} d\log W_{i_n}(X') \nonumber \\ 
& \, \times \, [W_{i_1},\ldots , W_{i_{n-1}}]_{X_0}(X')  \, ,
\end{align}
where the integration path $\gamma$ connects the boundary point $X_0$ to $X$. In the following we do not show explicitly the dependence on the kinematic point $X$.

In order to have a common notation, we rewrite the $\mathcal{Z}$ factor, as well as all other ingredients to the hard function, in the same iterated integral notation. In this way, we analytically perform simplifications for the hard function at the level of iterated integrals.
Remarkably, as observed previously for the planar case, we find that all weight three and four pieces cancel out.
Therefore we only need iterated integrals up to weight two. 
We rewrite them in terms of logarithms and dilogarithms.
For example,
\begin{align}
[ W_{1} ]_{X_{0}} &=  \log \left( s_{12}/3 \right) \,,\\
[ W_5 / W_2 , W_{12}/W_{2} ]_{X_{0}} &= -{\rm Li}_{2}\left( 1- {s_{15}}/{s_{23}} \right) \,.
\end{align}
Note that all functions are manifestly real-valued in the $s_{12}$-channel.
As a consequence, imaginary parts can only appear explicitly through the boundary values.

The analytic integrand expression, the IBP reductions, the $\epsilon$-expansion of the master integrals in terms of iterated integrals, as well as the infrared subtraction, are combined numerically using finite fields.  From this we reconstruct analytically the hard function.

At this stage, we make a remarkable observation: all dilogarithms and logarithms, as well as all imaginary parts,
can be absorbed into (the finite part of) one-mass box functions, which are defined as
\begin{align}
\label{eq:Ibox1m}
I_{123;45} = \ &  \text{Li}_2 \left( 1- {s_{12}}/{s_{45}}\right) + \text{Li}_2 \left( 1- {s_{23}}/{s_{45}}\right) \nonumber \\
& + \log^2 \left({s_{12}}/{s_{23}}\right) + {\pi^2}/{6} \, .
\end{align}
Considering all permutations of external momenta provides $30$ independent functions.
The analytic continuation of the latter to the physical scattering region is simply achieved by adding a small positive imaginary part to each two-particle Mandelstam invariant, $s_{ij} \rightarrow s_{ij} + i  0$.
This generates correctly all imaginary parts in the amplitude.

\sectioninline{Analytic result for the hard function.}
We express the hard function in the same coupling expansion (\ref{couplingexpansion}) and color decomposition \eqref{eq:ColorDecomposition} as the amplitude.
It can be written in terms of just two color components,
\begin{align}
\mathcal{H}^{(2)} = \sum_{S_5/S_{\mathcal{T}_1}}  \mathcal{T}_{1} \, \mathcal{H}^{(2)}_1 + \sum_{S_5/S_{\mathcal{T}_{13}}} \mathcal{T}_{13} \, \mathcal{H}^{(2)}_{13} \, ,
\end{align}
where each sum runs over the $5!$ permutations of the external legs, $S_5$, modulo the subset $S_{\mathcal{T}_{\lambda}}$ of permutations that leave $\mathcal{T}_{\lambda}$, and thus $\mathcal{H}^{(2)}_{\lambda}$, invariant.

Since the most sub-leading color components $\mathcal{H}^{(2,2)}_{\lambda}$ can be obtained from the planar $\mathcal{H}^{(2,0)}_{\lambda}$ and double trace $\mathcal{H}^{(2,1)}_{\lambda}$ ones through color relations~\cite{Edison:2011ta}, we present explicitly here only the latter:
\begin{widetext}
\begin{align}
\label{eq:H2T1}
\mathcal{H}^{(2,0)}_{1} = \ &
 \sum_{S_{\mathcal{T}_{1}}}  \Biggl\{ - 
 \kappa
  \,  \frac{[45]^2}{\vev{12} \vev{23} \vev{31}} \, I_{123;45} +
  \kappa^2
   \frac{1}{\vev{12} \vev{23} \vev{34} \vev{45} \vev{51}}  \biggl[ 5 \, s_{12} s_{23} + s_{12} s_{34} + \frac{\text{tr}_+^2(1245)}{s_{12} s_{45}} \biggr] \Biggr\} \, , \\
\label{eq:H2T13}
\mathcal{H}^{(2,1)}_{13} = \ & 
\sum_{S_{\mathcal{T}_{13}}}  \Biggl\{ 
 \kappa
 \, \frac{[15]^2}{\vev{23} \vev{34} \vev{42}} \, \biggl[  I_{234;15} + I_{243;15} - I_{324;15}  - 4 \, I_{345;12} - 4 \, I_{354;12} - 4 \, I_{435;12} \biggr]  \nonumber\\
& \hspace{1.5 cm} - 
6\, \kappa^2
 \biggl[ \frac{s_{23} \, \text{tr}_-(1345)}{s_{34} \,  \vev{12} \vev{23} \vev{34} \vev{45} \vev{51}} - \frac{3}{2} \frac{[12]^2}{\vev{34} \vev{45} \vev{53}} \biggr] \Biggr\}  \, ,
\end{align}
where $I$ was defined in Eq.~(\ref{eq:Ibox1m}), and  ${\rm tr}_{\pm}(ijkl) := \frac{1}{2} {\rm tr}[(1 \pm \gamma_{5}) \slashed{p}_i \slashed{p}_j  \slashed{p}_k \slashed{p}_l ]$.
\end{widetext}
The planar component~\eqref{eq:H2T1} is in agreement with the previous result in the literature~\cite{Gehrmann:2015bfy}.
The non-planar one~\eqref{eq:H2T13} is entirely new.
Remarkably, it exhibits the same striking simplicity: all functions of weight one, three, and four cancel out, and the remaining weight-two ones can all be expressed as permutations of the one-mass box function.
While the calculation was performed in the $s_{12}$ scattering region, the above formula can be
analytically continued to any other region by the $i0$ prescription mentioned above.

Note that the rational factors multiplying the transcendental part of the hard function~\eqref{eq:H2T1},~\eqref{eq:H2T13} are permutations of one object that appeared already in the one-loop amplitude~\eqref{A1loop}.
Remarkably, this object is conformally invariant. Moreover, the weight-two functions accompanying it are also governed by conformal symmetry. The latter manifests itself through anomalous conformal Ward identities \cite{Chicherin:2017bxc,Chicherin:2018ubl,Zoia:2018sin}.

\sectioninline{Verification of correct collinear factorization.}
In the limit where particles $1$ and $2$ are collinear, the full color five-gluon amplitude factorizes as follows:
\begin{align}
    \mathcal{A}^{(2)}&(1^+,2^+,3^+,4^+,5^+) \clim12 \nonumber\\
    &\mathcal{A}^{(2)}(P^+,3^+,4^+,5^+) \, \rm{Split}^{(0)}(-P^-;1^+,2^+) \nonumber\\
  + &\mathcal{A}^{(1)}(P^+,3^+,4^+,5^+) \, \rm{Split}^{(1)}(-P^-;1^+,2^+) \nonumber\\
  + &\mathcal{A}^{(1)}(P^-,3^+,4^+,5^+) \, \rm{Split}^{(1)}(-P^+;1^+,2^+) \, ,
\end{align}
where the sum goes over the color index of the gluon labelled by `$P$'. After inserting
expressions for the splitting amplitudes $\rm{Split}^{(\ell)}$~\cite{Parke:1986gb,Berends:1987me,Mangano:1987xk,Bern:1994zx} 
and four-gluon amplitudes~\cite{Bern:1991aq,Bern:2000dn},
we rewrite the collinear limit in terms of the trace decomposition (\ref{eq:ColorDecomposition}).
 
We verified the limits $\climj12$, $\climj23$ and $\climj34$ of the double trace term $\mathcal{T}_{13}$. 
It vanishes in the first two limits but has a non-trivial structure in the $\climj34$ limit. We find
perfect agreement.

\sectioninline{Discussion and outlook.}
In this Letter, we computed analytically, for the first time, all integrals needed for two-loop massless
five-particle scattering amplitudes in the physical scattering region. This required computing the master
integrals in all permutations of external legs, including their boundary values, in the physical scattering region.

In view of future phenomenological applications it is highly desirable to provide
fast numerical implementations of the non-planar pentagon functions computed here,
for example along the lines of  \cite{Gehrmann:2018yef}.

We used the expressions for the master integrals to compute analytically the five-gluon all-plus helicity amplitude at two loops.
The amplitude has the correct singularity structure and collinear behavior.
In the infrared-subtracted finite part, we observed remarkable cancellations of all weight one, three and four functions.

Intriguingly, we found that parts of the amplitude are governed by conformal symmetry.
It would be interesting to find an explanation for these observations.

Our work opens the door for further analytic calculations of massless two-loop five-particle amplitudes.
On the one hand, the complete information on the integral functions is now available.
On the other hand, the integral reductions required for the present calculation are of comparable complexity as
to what is expected to be needed for other helicity amplitudes, or amplitudes including fermions.

\sectioninline{Acknowledgments.}
We are grateful to T.\ Ahmed and B.\ Mistlberger for useful correspondence,
and  to S.\ Jahn for useful discussions. We thank the HPC groups at JGU Mainz and MPCDF for support.
S.\ B. is supported by an STFC Rutherford Fellowship ST/L004925/1.
This research received funding from Swiss National Science Foundation (Ambizione grant PZ00P2 161341),
the European Research Council (ERC) under the European Union's Horizon 2020 research and innovation programme, {\it Novel
structures in scattering amplitudes} (grant agreement No 725110), and {\it High precision multi-jet dynamics at the LHC} (grant agreement No 772009), under the Marie Skłodowska-Curie grant agreement 746223,
and from the COST Action CA16201 Particleface.

\bibliographystyle{h-physrev5}

\bibliography{5point_refs4}

\begin{thebibliography}{10}

\bibitem{Aaboud:2017fml}
ATLAS, M.~Aaboud {\em et~al.},
\newblock Eur. Phys. J. {\bf C77}, 872 (2017), arXiv:1707.02562.

\bibitem{Sirunyan:2018adt}
CMS, A.~M. Sirunyan {\em et~al.},
\newblock JHEP {\bf 12}, 117 (2018), arXiv:1811.00588.

\bibitem{Currie:2017eqf}
J.~Currie {\em et~al.},
\newblock Phys. Rev. Lett. {\bf 119}, 152001 (2017), arXiv:1705.10271.

\bibitem{Bern:1993mq}
Z.~Bern, L.~J. Dixon, and D.~A. Kosower,
\newblock Phys. Rev. Lett. {\bf 70}, 2677 (1993), arXiv:hep-ph/9302280.

\bibitem{Bern:1994fz}
Z.~Bern, L.~J. Dixon, and D.~A. Kosower,
\newblock Nucl. Phys. {\bf B437}, 259 (1995), arXiv:hep-ph/9409393.

\bibitem{Kunszt:1994nq}
Z.~Kunszt, A.~Signer, and Z.~Trocsanyi,
\newblock Phys. Lett. {\bf B336}, 529 (1994), arXiv:hep-ph/9405386.

\bibitem{vonManteuffel:2014ixa}
A.~von Manteuffel and R.~M. Schabinger,
\newblock Phys. Lett. {\bf B744}, 101 (2015), arXiv:1406.4513.

\bibitem{Larsen:2015ped}
K.~J. Larsen and Y.~Zhang,
\newblock Phys. Rev. {\bf D93}, 041701 (2016), arXiv:1511.01071.

\bibitem{Peraro:2016wsq}
T.~Peraro,
\newblock JHEP {\bf 12}, 030 (2016), arXiv:1608.01902.

\bibitem{Badger:2017jhb}
S.~Badger, C.~Brønnum-Hansen, H.~B. Hartanto, and T.~Peraro,
\newblock Phys. Rev. Lett. {\bf 120}, 092001 (2018), arXiv:1712.02229.

\bibitem{Boehm:2017wjc}
J.~Boehm, A.~Georgoudis, K.~J. Larsen, M.~Schulze, and Y.~Zhang,
\newblock Phys. Rev. {\bf D98}, 025023 (2018), arXiv:1712.09737.

\bibitem{Boehm:2018fpv}
J.~Boehm, A.~Georgoudis, K.~J. Larsen, H.~Schönemann, and Y.~Zhang,
\newblock JHEP {\bf 09}, 024 (2018), arXiv:1805.01873.

\bibitem{Kosower:2018obg}
D.~A. Kosower,
\newblock Phys. Rev. {\bf D98}, 025008 (2018), arXiv:1804.00131.

\bibitem{Chawdhry:2018awn}
H.~A. Chawdhry, M.~A. Lim, and A.~Mitov,
\newblock Phys. Rev. {\bf D99}, 076011 (2019), arXiv:1805.09182.

\bibitem{Ita:2015tya}
H.~Ita,
\newblock Phys. Rev. {\bf D94}, 116015 (2016), arXiv:1510.05626.

\bibitem{Abreu:2017idw}
S.~Abreu, F.~Febres~Cordero, H.~Ita, M.~Jaquier, and B.~Page,
\newblock Phys. Rev. {\bf D95}, 096011 (2017), arXiv:1703.05255.

\bibitem{Gehrmann:2015bfy}
T.~Gehrmann, J.~M. Henn, and N.~A. Lo~Presti,
\newblock Phys. Rev. Lett. {\bf 116}, 062001 (2016), arXiv:1511.05409,
\newblock [Erratum: Phys. Rev. Lett.116,no.18,189903(2016)].

\bibitem{Papadopoulos:2015jft}
C.~G. Papadopoulos, D.~Tommasini, and C.~Wever,
\newblock JHEP {\bf 04}, 078 (2016), arXiv:1511.09404.

\bibitem{Gehrmann:2018yef}
T.~Gehrmann, J.~M. Henn, and N.~A. Lo~Presti,
\newblock JHEP {\bf 10}, 103 (2018), arXiv:1807.09812.

\bibitem{Chicherin:2018mue}
D.~Chicherin {\em et~al.},
\newblock JHEP {\bf 03}, 042 (2019), arXiv:1809.06240.

\bibitem{Chicherin:2018old}
D.~Chicherin {\em et~al.},
\newblock (2018), arXiv:1812.11160.

\bibitem{Abreu:2018aqd}
S.~Abreu, L.~J. Dixon, E.~Herrmann, B.~Page, and M.~Zeng,
\newblock Phys. Rev. Lett. {\bf 122}, 121603 (2019), arXiv:1812.08941.

\bibitem{Chicherin:2018yne}
D.~Chicherin {\em et~al.},
\newblock Phys. Rev. Lett. {\bf 122}, 121602 (2019), arXiv:1812.11057.

\bibitem{Chicherin:2019xeg}
D.~Chicherin {\em et~al.},
\newblock JHEP {\bf 03}, 115 (2019), arXiv:1901.05932.

\bibitem{Abreu:2019rpt}
S.~Abreu, L.~J. Dixon, E.~Herrmann, B.~Page, and M.~Zeng,
\newblock JHEP {\bf 03}, 123 (2019), arXiv:1901.08563.

\bibitem{Badger:2013gxa}
S.~Badger, H.~Frellesvig, and Y.~Zhang,
\newblock JHEP {\bf 12}, 045 (2013), arXiv:1310.1051.

\bibitem{Abreu:2017hqn}
S.~Abreu, F.~Febres~Cordero, H.~Ita, B.~Page, and M.~Zeng,
\newblock Phys. Rev. {\bf D97}, 116014 (2018), arXiv:1712.03946.

\bibitem{Abreu:2018jgq}
S.~Abreu, F.~Febres~Cordero, H.~Ita, B.~Page, and V.~Sotnikov,
\newblock JHEP {\bf 11}, 116 (2018), arXiv:1809.09067.

\bibitem{Badger:2018enw}
S.~Badger, C.~Brønnum-Hansen, H.~B. Hartanto, and T.~Peraro,
\newblock JHEP {\bf 01}, 186 (2019), arXiv:1811.11699.

\bibitem{Abreu:2018zmy}
S.~Abreu, J.~Dormans, F.~Febres~Cordero, H.~Ita, and B.~Page,
\newblock Phys. Rev. Lett. {\bf 122}, 082002 (2019), arXiv:1812.04586.

\bibitem{Abreu:2019odu}
S.~Abreu {\em et~al.},
\newblock (2019), arXiv:1904.00945.

\bibitem{Badger:2015lda}
S.~Badger, G.~Mogull, A.~Ochirov, and D.~O'Connell,
\newblock JHEP {\bf 10}, 064 (2015), arXiv:1507.08797.

\bibitem{tHooft:1972tcz}
G.~'t~Hooft and M.~J.~G. Veltman,
\newblock Nucl. Phys. {\bf B44}, 189 (1972).

\bibitem{Bern:2002zk}
Z.~Bern, A.~De~Freitas, L.~J. Dixon, and H.~L. Wong,
\newblock Phys. Rev. {\bf D66}, 085002 (2002), arXiv:hep-ph/0202271.

\bibitem{Parke:1985pn}
S.~J. Parke and T.~R. Taylor,
\newblock Phys. Lett. {\bf 157B}, 81 (1985),
\newblock [Erratum: Phys. Lett.B174,465(1986)].

\bibitem{Mangano:1990by}
M.~L. Mangano and S.~J. Parke,
\newblock Phys. Rept. {\bf 200}, 301 (1991), arXiv:hep-th/0509223.

\bibitem{Edison:2011ta}
A.~C. Edison and S.~G. Naculich,
\newblock Nucl. Phys. {\bf B858}, 488 (2012), arXiv:1111.3821.

\bibitem{Witten:2003nn}
E.~Witten,
\newblock Commun. Math. Phys. {\bf 252}, 189 (2004), arXiv:hep-th/0312171.

\bibitem{Dunbar:2016aux}
D.~C. Dunbar and W.~B. Perkins,
\newblock Phys. Rev. {\bf D93}, 085029 (2016), arXiv:1603.07514.

\bibitem{Catani:1998bh}
S.~Catani,
\newblock Phys. Lett. {\bf B427}, 161 (1998), arXiv:hep-ph/9802439.

\bibitem{Becher:2009cu}
T.~Becher and M.~Neubert,
\newblock Phys. Rev. Lett. {\bf 102}, 162001 (2009), arXiv:0901.0722,
\newblock [Erratum: Phys. Rev. Lett.111,no.19,199905(2013)].

\bibitem{Gardi:2009qi}
E.~Gardi and L.~Magnea,
\newblock JHEP {\bf 03}, 079 (2009), arXiv:0901.1091.

\bibitem{Lee:2012cn}
R.~N. Lee,
\newblock (2012), arXiv:1212.2685.

\bibitem{ArkaniHamed:2010gh}
N.~Arkani-Hamed, J.~L. Bourjaily, F.~Cachazo, and J.~Trnka,
\newblock JHEP {\bf 06}, 125 (2012), arXiv:1012.6032.

\bibitem{Henn:2013pwa}
J.~M. Henn,
\newblock Phys. Rev. Lett. {\bf 110}, 251601 (2013), arXiv:1304.1806.

\bibitem{Henn:2019rmi}
J.~M. Henn, T.~Peraro, M.~Stahlhofen, and P.~Wasser,
\newblock (2019), arXiv:1901.03693.

\bibitem{Henn:2019rgj}
J.~M. Henn and B.~Mistlberger,
\newblock (2019), arXiv:1902.07221.

\bibitem{Chicherin:2017dob}
D.~Chicherin, J.~Henn, and V.~Mitev,
\newblock JHEP {\bf 05}, 164 (2018), arXiv:1712.09610.

\bibitem{Gehrmann:2000zt}
T.~Gehrmann and E.~Remiddi,
\newblock Nucl. Phys. {\bf B601}, 248 (2001), arXiv:hep-ph/0008287.

\bibitem{Gehrmann:2001ck}
T.~Gehrmann and E.~Remiddi,
\newblock Nucl. Phys. {\bf B601}, 287 (2001), arXiv:hep-ph/0101124.

\bibitem{Borowka:2017idc}
S.~Borowka {\em et~al.},
\newblock Comput. Phys. Commun. {\bf 222}, 313 (2018), arXiv:1703.09692.

\bibitem{Borowka:2018goh}
S.~Borowka {\em et~al.},
\newblock Comp. Phys. Comm.  (2018), arXiv:1811.11720.

\bibitem{Chicherin:2017bxc}
D.~Chicherin and E.~Sokatchev,
\newblock JHEP {\bf 04}, 082 (2018), arXiv:1709.03511.

\bibitem{Chicherin:2018ubl}
D.~Chicherin, J.~M. Henn, and E.~Sokatchev,
\newblock Phys. Rev. Lett. {\bf 121}, 021602 (2018), arXiv:1804.03571.

\bibitem{Zoia:2018sin}
S.~Zoia,
\newblock PoS {\bf LL2018}, 037 (2018), arXiv:1807.06020.

\bibitem{Parke:1986gb}
S.~J. Parke and T.~R. Taylor,
\newblock Phys. Rev. Lett. {\bf 56}, 2459 (1986).

\bibitem{Berends:1987me}
F.~A. Berends and W.~T. Giele,
\newblock Nucl. Phys. {\bf B306}, 759 (1988).

\bibitem{Mangano:1987xk}
M.~L. Mangano, S.~J. Parke, and Z.~Xu,
\newblock Nucl. Phys. {\bf B298}, 653 (1988).

\bibitem{Bern:1994zx}
Z.~Bern, L.~J. Dixon, D.~C. Dunbar, and D.~A. Kosower,
\newblock Nucl. Phys. {\bf B425}, 217 (1994), arXiv:hep-ph/9403226.

\bibitem{Bern:1991aq}
Z.~Bern and D.~A. Kosower,
\newblock Nucl. Phys. {\bf B379}, 451 (1992).

\bibitem{Bern:2000dn}
Z.~Bern, L.~J. Dixon, and D.~A. Kosower,
\newblock JHEP {\bf 01}, 027 (2000), arXiv:hep-ph/0001001.

\end{thebibliography}

\end{document}